\documentclass{article}
\usepackage{icrctc07}

\title{3D Reconstruction of Extensive Air Showers from Fluorescence Data}
\shorttitle{3D Reconstruction of EAS}
\authors{S. Andringa$^1$, M. Pato$^1$ and M. Pimenta$^1$, for the Pierre Auger Collaboration}
\shortauthors{S. Andringa and et al}
\afiliations{$^1$Laborat\'orio de Instrumenta\c{c}\~ao e F\'\i sica de Part\'\i culas, \\Av. Elias Garcia, 14, 1$^o$, 1000-149 Lisboa, Portugal}
\email{sofia@lip.pt}

\abstract{
A new method to reconstruct the 3-dimensional structure of extensive 
air showers, seen by fluorescence detectors, is proposed.
The observation of the shower is done in 2-dimensional pixels, 
for consecutive time bins. Time corresponds to a third dimension.
Assuming that the cosmic ray shower propagates as a plane wave 
front moving at the speed of light, a complex 3D volume in space
can be associated to each measured charge (per pixel and time bin). 
The 3D description in space allows a simultaneous access to the 
longitudinal and lateral profiles of each shower. In the case that 
several eyes observe the same shower, the method gives a straight-forward 
combination of all the information. This method is in an early phase
of development and is not used for the general reconstruction of the Auger data. 
}
\begin{document}
\maketitle
%\linenumbers
%Begin the section.
\section{Introduction}
The Pierre Auger Observatory will provide a large set of cosmic ray data 
to be analysed in multiple perspectives, ranging from particle physics to
cosmology. A detailed understanding of the data is crucial. 

The method proposed in this contribution aims at reconstructing fine-details
of the individual shower structure and being sensitive to both lateral and 
longitudinal profiles simultaneously, from the fluorescence light emission. 
It is not the standard reconstruction method used in Auger, but is built on
top of the existing methods that provide the geometry of the shower line and
of the longitudinal profiles with great accuracy.

The Auger Fluorescence Detector (FD) is composed of four eyes, each with six 
telescopes, observing the atmosphere over the centre of the Surface Detector 
array from different directions, with each eye covering elevations 
2$^\circ$ to 32$^\circ$ and 180$^\circ$ in azimuth. The data comprises cosmic 
ray events up to the highest energies, at very different distances from the 
detector eyes - ranging from 1 to above 30 km. 

The Auger FD data are collected in pixels of approximate 
angular dimensions of 1.5$^\circ$ and in 100 ns time bins. 
The standard reconstruction is based first on the pixel angular information,
to find the plane that contains the shower axis and the observing eye (the 
Shower-Detector Plane, SDP). The centroid time found for each pixel is then 
used to find the axis line within this plane, the minimum approach distance 
to the detector, and the reference time at which it occurred ($T_0$).
The shower geometry is simply given by this line
(details of the calculation are given in \cite{standard_rec}, the reconstruction
can also use several eyes and the surface detector, as explained in \cite{hybrid_rec}). 
In the above procedure pixels which do not observe the SDP cannot be used for the 
geometry reconstruction and the time structure of the signal inside each pixel is neglected. 
Although for distant showers the line approach is clearly sufficient, for 
close-by showers relevant information can be lost.

To measure the energy deposited by the primary cosmic ray, a Gaisser-Hillas 
function is then fitted on this line. To include the effect of non-perfect 
optics and the fact that the shower in not only a line, the contribution of 
near-by pixels is considered at this stage, by summing signals for each time
bin. Some of the information important for the longitudinal profile and energy 
reconstruction is restored in this procedure.
% -- however, since
%the same time of observation in different positions corresponds to different 
%emission times, the treatment is not exact.

However, the information that can be present in the lateral profile of the 
shower, in possible asymmetries or local fine structures is lost in the 
standard reconstruction -- this is the main motivation to find another 
reconstruction method that takes all the available information into account.
The method described below is under test in simulation (and laser data), 
and will soon be tested in real data.

\section{The 3D method}
\subsection{Geometry Reconstruction}

The basic idea of the method is to propagate the available 3D information - binned
in two angular dimensions and one time dimension - into 3D volumes in space. These
volumes contain the points from which the observed fluorescence photons were emitted.

The cosmic ray shower is assumed to propagate as a plane wave front, moving at
the speed of light. For each direction observed in the detector, there is a line
in space that in general does not intercept the shower axis. The observation time, 
on the other hand, is a sum of the emission time of the photon 
and its propagation time to the detector. The emission time is 
characteristic of all the particles in a disk moving coherently along the shower 
axis at constant velocity c.

Starting from a given hypothesis for the shower axis and core distance, each 
observation direction at each time corresponds to a unique point in space.
The axis and core coming from the standard reconstruction (being it mono, stereo 
or hybrid when available) are used as a first hypothesis, and, with the geometry fixed, 
the time of closest approach between the shower line and the detector, $T_0$, is found 
using the pixel centroids and directions within the SDP, as before.

The position of the centre of each volume (corresponding to $\theta_j, \phi_j$
of the pixel centre and $t_i$ for each time bin) is then found for 
\begin{equation}\label{equ1}
r_{ij}=\frac{c\cdot(t_i-T_0)}{(1+cos(\alpha_j))}, 
\end{equation}
being $\alpha_j$ the angle between the observation direction and the shower axis.

The borders of each volume are given by the same procedures using the $\theta,\phi$
of each vertex and $t_i \pm 50$ ns. Notice that the Auger pixels are regular hexagons
layed on a spherical surface -- irregular in $\theta,\phi$ -- and that the observation 
times correspond to constant emission times along the bisectrix between the shower axis 
and the observation line.

\begin{figure}[t]
\begin{center}
\includegraphics[width=.48\textwidth]{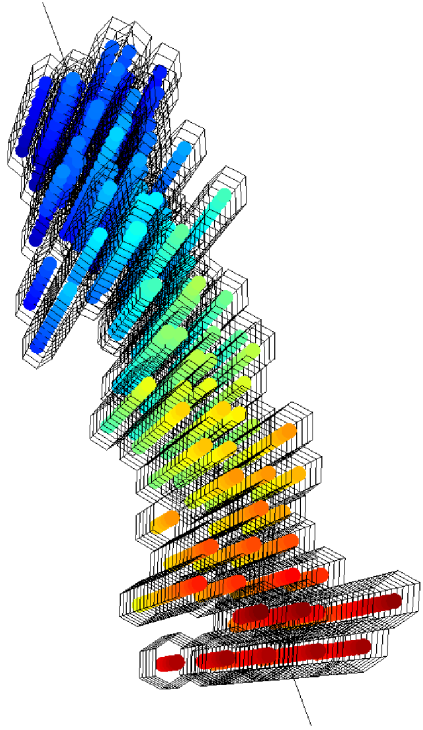}
\caption{\label{fig:display} 3D view of a simulated event %\cite{SIM}  
of 10$^{18.5}$~eV seen at 2 km from the telescope.
The shapes of the 3D volumes are shown and
the color code corresponds to the detection time.
The visualisation is done with the map3d package \cite{3dViewer}.}
\end{center}
\end{figure}

As the time differences between pixels give information on the shower axis location,
the time duration of the signal within each pixel can also provide the same information.
In addition, larger time signals can also correspond to larger lateral dimensions of the 
shower. To profit from the extra information, the charge observed in each volume is used 
to refine the shower axis (and core) determination.

This is done by calculating the ``centre-of-mass'' and ``inertia axis'' of the 3D object 
composed by all the volumes, considering the charge to be, in first approximation, proportional
the number of particles, and to the ``mass''.

The main ``inertia axis'' gives the shower direction, replacing the shower axis (the shower 
core being fixed by the ``centre-of-mass'').The two other axes give a first measurement of the
average lateral distribution, the fact that they are usually equal shows that there is no 
significant bias in the use of time as a third dimension.

To a new axis and core hypothesis correspond new SDP and $T_0$ (the latest to be recalculated 
from data), and so new positions and shapes for each volume. 
The procedure can thus be iterated until reasonable stability is obtained for the axis direction 
and core location. The resolutions on core location and axis directions obtained after the iterations 
are comparable with the ones from the standard reconstruction used as inputs, no clear improvement is
obtained but no information is lost, which again shows that there is no 
significant bias in the use of time as a third dimension.

% A. Watson: some numbers for geometry, dTh and DCore...

Figure \ref{fig:display} shows the 3D view of a simulated close-by event. The different volume
shapes with dimensions coming from pixel and time bin size are clearly seen, the color codes
show the time bin of detection associated to each volume.

\subsection{Energy and Profiles Reconstruction}

The 3D volumes in space correspond to the regions from which the observed photons were emitted.
It is then possible to compare their charges to that given by a hypothesis - a longitudinal and
a lateral profile - taking into account the light attenuation from the emission point ($P$) to 
the detector and the effective collection area seen by the telescope, as shown in equation
\ref{eq:theory}.

\begin{eqnarray}\label{eq:theory}
%Exp(P) = Y_f \times GH(X_P) \times GO(X_P,R_P) \times RS(r_P) \times A\cos{\theta_P}/(4\pi r_P^2)
%Exp_P=LongP_{Fl}(X_P)\cdot LatP_{Fl}(X_P,R_P)\cdot RS(r_P)\cdot\\
Exp_P=GH({\bf X_P},X_{max},X_0,\lambda,N_{max})\cdot\\
\nonumber  GO({\bf X_P,R_P},X_{max})\cdot RS({\bf r_P})\cdot\\
\nonumber Y_f\cdot\frac{A}{(4\pi {\bf r_P}^2)}
\end{eqnarray}

where $Y_f$ is the fluorescence yield, $GH$ is the longitudinal profile depending on the slant 
depth, $X_P$, and $GO$ is the lateral profile, that depends also on $R_P$ the distance to the axis 
(the profiles are given, respectively, by a Gaisser-Hillas and a G\'ora function \cite{GORAfunc}, 
for example),
$RS$ is the Rayleigh Scattering factor, depending on the distance between the 
emission and detection points $r_P$ (no Mie scattering is considered for now and 
multiple scattering contributions are neglected), and $A$ represents the effective
diaphragm area and may include constant calibration factors. 

In addition to the fluorescence light, also direct and scattered Cherenkov are included.
Even for showers not directed to the eye, scattered Cherenkov can represent a
non-negligible fraction of the detected light.
The parameterisation of the angular distribution function given in ref~\cite{nerling} is used
for the direct contribution, while for the scattered one the convolution of the Cherenkov 
light production with Rayleigh scattering is considered.

The light observed in the telescope is spread by the non-perfect optics in the mirror, producing a
spot, which depends on the incidence angle and can be parameterised from simulation, including
also camera shadow effects. This information can be included to find the exact positions of the photons 
in the camera. Since the calibration is done by illuminating equally all the camera 
points, a factor $f_m$ can be found to correct up the sensitive areas, and correct 
down the Mercedes regions. 
%\cite{SPOT}.

The Monte Carlo integration starts by finding a cylinder with the reconstructed shower axis, and
enclosing the several reconstructed volumes.
The expected number of photons from function \ref{eq:theory} is evaluated for each of 
$N_1 \times N_{vol}$ randomly generated points, to cover all the $N_{vol}$ volumes found for 
each telescope. Afterwards,
$N_2$ points are spread according to the spot parametrisation and, in each, the function is
corrected by $f_m$. The sum of the function for the final points in each volume, normalised by 
volume of the cylinder $V_{cyl}$, gives the expected number of photons per pixel and time bin, 
to be compared to the observed one:

\begin{equation}\label{fun:exp}
Exp_{vol} = \frac{V_{cyl}}{N_{vol}N_1N_2} \sum_{P}{Exp_P\sum_{P' \in vol} f_m(P')}
\end{equation}

\begin{figure}[h]
\begin{center}
\includegraphics[angle=270., width=.50\textwidth]{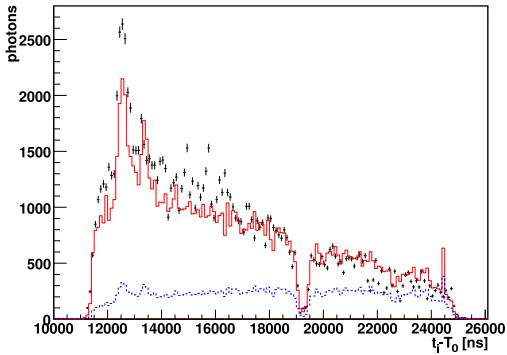}
\end{center}
\caption{\label {fig:charge} Number of observed photons from a simulated event
of 10$^{18.5}$~eV, as a function of time. The data points are compared to the 
total expectations. The Cherenkov contribution is shown separately in the dashed line.}
\end{figure}

Figure \ref{fig:charge} shows the expected and observed number of detected photons as a function
of the observed time, for a given simulated event.
The simulated values are used for the Gaisser-Hillas and G\'ora function. Knowing the error on the 
observed value, one can create a $\chi^2$ function to minimise for the model parameters appearing 
in both functions. The lateral profile function could be directly tested for close-by events and
%as pointed out in \cite{Gora_Xmax}, 
that might help in the estimation of the total energy and 
$X_{max}$, even when $X_{max}$ is out of the field of view.

\nocite{standard_rec}
\nocite{hybrid_rec}
\nocite{3dViewer}
\nocite{GORAfunc}
%This is the reference to .bib file (Whitout .bib!)
\bibliography{3drec_icrc07}
%This in the bibtex style, is ok.
\bibliographystyle{plain}

\end{document}